\documentclass[aps,pra,reprint,superscriptaddress,floatfix]{revtex4-2}

\usepackage[colorlinks=true,urlcolor=blue]{hyperref}
\usepackage{amsmath,amssymb,bm}
\usepackage{graphicx}
\usepackage{float}
\usepackage{cancel}
\usepackage{physics}

\usepackage{tcolorbox}

\begin{document}

\title{Supersolid formation in a dipolar condensate by roton instability}

\author{Aitor Ala\~na}
\affiliation{Department of Physics, University of the Basque Country UPV/EHU, 48080 Bilbao, Spain}
\affiliation{EHU Quantum Center, University of the Basque Country UPV/EHU, 48940 Leioa, Biscay, Spain}

\author{I\~nigo L. Egusquiza}
\affiliation{Department of Physics, University of the Basque Country UPV/EHU, 48080 Bilbao, Spain}
\affiliation{EHU Quantum Center, University of the Basque Country UPV/EHU, 48940 Leioa, Biscay, Spain}

\author{Michele Modugno}
\affiliation{Department of Physics, University of the Basque Country UPV/EHU, 48080 Bilbao, Spain}
\affiliation{EHU Quantum Center, University of the Basque Country UPV/EHU, 48940 Leioa, Biscay, Spain}
\affiliation{IKERBASQUE, Basque Foundation for Science, 48013 Bilbao, Spain}

\date{\today}

\begin{abstract}
We characterize the role of the roton instability in the formation of a supersolid state of an elongated dipolar condensate, following a quench of the contact interactions across the superfluid-supersolid transition, as observed in recent experiments. We perform dynamical simulations by means of the extended Gross-Pitaevskii equation including quantum corrections, for different final values of the $s-$wave scattering length. The corresponding excitation spectrum is computed using an effective one-dimensional description, revealing that the calculated growth rates of the unstable roton mode accurately reproduce the observed behavior. Our results provide valuable insights regarding the formation time of the supersolid and its scaling behavior with respect to the $s$-wave scattering length.
\end{abstract}

\maketitle

\section{Introduction}

Ultracold dipolar gases, characterized by significant magnetic or electric dipole moments, have emerged as a fascinating area of research in the field of cold atoms and quantum gases. Unlike traditional Bose-Einstein condensates (BECs), with only short-range contact interactions, dipolar condensates exhibit also long-range, anisotropic interactions (see, e.g., Refs. \cite{lahaye2009,baranov2012}). Notably, their excitation spectrum is characterized by a rotonic mode \cite{Sa03,odell2003,ronen2007,Ch18,Pe19,Sc21}, which may become unstable \cite{giovanazzi2002,Ro19} and lead to exotic and complex behavior, including the formation of droplets \cite{kadau2016} and supersolids \cite{Ta19,Bo19,Ch19}, among others \cite{wenzel2017,klaus2022}.

In particular, in recent years an intense research activity has been directed towards unraveling the properties of the supersolid phase of dipolar gases, both from the experimental \cite{pollet2019,donner2019,Ta19b,Gu19,Na19,Ta21,He21,Pe21,No21,bottcher2021,sohmen2021,biagioni2022,Bl22,sanchez-baena2022} and theoretical point of view
\cite{Zh19,schuster20,blakie2020variational,Ro20,gallemi2020,Bl20,Te21,Zh21,He21b,He21c,alana2022crossing,roccuzzo2022,ilg2023,smith2023}. 
Proposed in the past century \cite{Le70,chester70}, supersolids are an exciting phase of matter combining a superfluid nature \cite{Gross62,pitaevskii2016} with the translational symmetry breaking characteristic of solid structures \cite{Gross57,cristSov70,boni12,yukalov2020}. 
Notably, the supersolid (SS) phase of matter can be achieved through both a classical transition from a gas to a supersolid \cite{sohmen2021} and a quantum transition from an unmodulated superfluid (SF) to a supersolid. 
Experimentally and theoretically both discontinuous \cite{Po94,Ma13,Lu15,Ta19,Bo19,Ta19b} and continuous \cite{Se08,Pe21} features have been observed, reminiscent of the first-order and second-order transitions predicted in the thermodynamic limit in two dimensions (2D) and one dimension (1D), respectively. 
Interestingly, the effective dimensionality of the system can be controlled by tuning the transverse confinement and the atom number, as experimentally observed by Biagioni \textit{et al.} \cite{biagioni2022} and theoretically discussed in Ref. \cite{alana2022crossing}. 

A remarkable feature of this transition is that whereas one can relax an initial SS state onto a SF state by crossing the transition almost adiabatically, in the opposite direction the supersolid requires a finite time to form \cite{Bo19,biagioni2022,alana2022crossing,He21}. This corresponds to the time required for intrinsic fluctuations -- associated to the roton instability -- to grow up and break the translational invariance of the system.

Here, we provide a theoretical characterization of the role of this instability in the  dynamics of formation of a supersolid in a quasi-1D dipolar condensate, by considering a typical configuration of the experiment in Ref. \cite{biagioni2022}. In particular, we focus on the scenario in which the system crosses the SF-SS phase transition by undergoing a quench of the contact interactions which, for quasi-1D systems, can still provide a smooth connection between an unmodulated BEC and a supersolid array \cite{alana2022crossing} (unlike what occurs in 2D \cite{Bl22}).
We perform a systematic analysis by means of numerical simulations of the extended Gross-Pitaevskii equation, considering various amplitudes of the interaction quench. The corresponding roton spectrum is computed by means of the 1D theory proposed by Blakie \textit{et al.} \cite{blakie2020variational}, which provides a simplified yet accurate characterization of the simulation outcomes.
These results are then used to discuss the general features that determine the formation time of the supersolid and its scaling behavior with respect to the $s$-wave scattering length.

The paper is organized as follows. In Section \ref{sec:system}, we provide an overview of the system under consideration and briefly summarize the relevant formulas defining the extended Gross-Pitaevskii theory for dipolar condensates. In Sec. \ref{sec:protocol}, we discuss the general protocol employed to induce the interaction quench across the SF-SS phase transition, presenting the general phenomenology observed through numerical simulations. In Sec. \ref{sec:1Dmodel} we briefly review the 1D effective theory of Ref. \cite{blakie2020variational} used to study the excitation spectrum of an elongated dipolar condensate. The latter is computed for our specific case in Sec. \ref{sec:stability}, where it is used to characterize systematically the role of the roton instability in the formation of the supersolid. There, we also explore the implications of this instability on the scaling behavior of the supersolid formation time with respect to the $s$-wave scattering length. Finally, we present a summary of our findings and concluding remarks in Sec. \ref{sec:conclusions}.

\section{System}
\label{sec:system}

In the following analysis, we consider the quasi 1D configuration investigated in the experiment of Ref. \cite{biagioni2022} and theoretically analyzed in Ref. \cite{alana2022crossing}.
It consists of a dipolar condensate, at zero temperature, 
composed by $N=3\times10^{4}$ magnetic atoms of $^{162}$Dy -- with tunable $s$-wave scattering length $a_{s}$ and dipolar length $a_{dd}=130a_0$ ($a_0$ being the Bohr radius) -- trapped by a harmonic potential with frequencies $(\omega_{x},\omega_{y},\omega_{z})=2\pi\times(15, 101, 94)$ Hz. While the choice of these parameters is motivated by their experimental feasibility in a specific case, it is worth noting that the following analysis is conceptually general, which enables its extension to other scenarios.

This system can be described in terms of an extended Gross-Pitaevskii (GP) theory including dipolar interactions \cite{ronen2006} and the Lee-Huang-Yang (LHY) correction accounting for quantum fluctuations, within the local density approximation \cite{Li12,wachtler2016,schmitt2016}. The energy functional can be written as $E = E_{\text{GP}} + E_{\text{dd}} + E_{\text{LHY}}$ with
\begin{align}
E_{\text{GP}} &= 
\int \left[\frac{\hbar^2}{2m}|\nabla \psi(\bm{r})|^2 + V(\bm{r})n(\bm{r})+\frac{g}{2} n^2(\bm{r})
\right]d\bm{r}\,,
\nonumber\\
E_{\text{dd}} &=\frac{C_{\text{dd}}}{2}\iint n(\bm{r})V_{\text{dd}}(\bm{r}-\bm{r}')n(\bm{r}') d\bm{r}d\bm{r}'\,,
\label{eq:GPenergy}\\
E_{\text{LHY}} &=\frac{2}{5}\gamma_{\text{LHY}}\int n^{5/2}(\bm{r})d\bm{r}\,,
\nonumber
\end{align}
where $E_{\text{GP}}=E_{\text{k}}+E_{\text{ho}}+E_{\text{int}}$ is the standard GP energy functional including the kinetic, potential, and contact interaction terms, $V(\bm{r})=(m/2)\sum_{\alpha=x,y,z}\omega_{\alpha}^{2}r_{\alpha}^{2}$ is the harmonic trapping potential, $n(\bm{r})=|\psi(\bm{r})|^2$ represents the condensate density (normalized to the total number of atoms $N$), $g=4\pi\hbar^2 a_{s}/m$ is the contact interaction strength, $V_{\text{dd}}(\bm{r})= (1-3\cos^{2}\theta)/(4\pi r^{3})$ the inter-particle dipole-dipole potential, $C_{\text{dd}}\equiv\mu_{0}\mu^2$ its strength, $\mu$ the modulus of the dipole moment $\bm{\mu}$, $\bm{r}$ the distance between the dipoles, and $\theta$ the angle between the vector $\bm{r}$ and the dipole axis, $\cos\theta=\bm{\mu}\cdot\bm{r}/(\mu r)$. 
As in Refs. \cite{biagioni2022,alana2022crossing}, we consider the magnetic dipoles to be aligned along the $z$ direction by a magnetic field $\bm{B}$.
The LHY coefficient is $\gamma_{\text{LHY}}={128\sqrt{\pi}}{\hbar^{2}a_s^{5/2}}/(3m)\left(1 + 3\epsilon_{\text{dd}}^{2}/2\right)$, with $\epsilon_{\text{dd}}=\mu_0 \mu^2 N/(3g)$.

As discussed in Refs. \cite{biagioni2022,alana2022crossing}, the equilibrium configuration of the system corresponds to either a conventional superfluid (SF) state or a supersolid (SS) state. The transition from one phase to the other can be induced by tuning the $s$-wave scattering length $a_{s}$. For the present values of the number of atoms and trapping frequencies, the critical point is located at $a_{s}^{c} \simeq 94.4 a_{0}$ and the transition has a continuous character \cite{biagioni2022,alana2022crossing}.

\section{Protocol}
\label{sec:protocol}

In order to study the formation of a supersolid, we adopt the following approach. The system is initially prepared in an equilibrium configuration within the SF phase, at $a_s^{(in)} = a_s^c + 1.5\,a_{0}$. For the sake of simplicity, this value is kept fixed throughout this work. Then, the condensate is quenched into the SS phase by a sudden change of the contact scattering length, to a final value $a_s^{(fi)} = a_s^c - \delta a_{s}$ . 
In the present study, $\delta a_{s}$ varied in the range $[1.0,6.5]a_{0}$.

The dynamics of the system following the quench is obtained by solving the GP equation \cite{pitaevskii2016}
\begin{equation}
 \label{eq:gp}
 i\hbar{\partial_{t}\psi}={\delta E[\psi,\psi^*]}/{\delta \psi^*},
\end{equation}
where the energy functional $E[\psi,\psi^*]$ is the one in Eq. (\ref{eq:GPenergy}) \footnote{For the details of the numerical methods see Ref. \cite{alana2022crossing}.}.
To uncover the physics involved in the formation process, we simplify the analysis by excluding dissipation mechanisms and particle losses, which are not essential for the present discussion.

\begin{figure}[t]
	\centerline{\includegraphics[width=\linewidth]{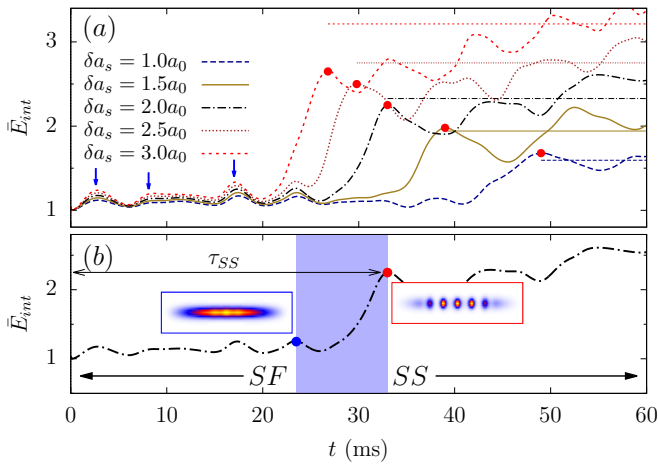}}
 \caption{(a) Behavior of $\bar{E}_{\text{int}}(t)$ during the evolution after the quench, for different values of $\delta a_s$ ranging from $1\,a_0$ to $3\,a_0$.
 The red dots correspond to the values $\bar{E}_{\text{int}}^*$ at which the exponential growth saturates; the blue arrows the oscillations maxima before the formation of the SS pattern. The horizontal lines indicate the equilibrium value of $\bar{E}_{\text{int}}^{SS}$ of the SS ground state at the same value of $\delta a_s$ (same line type).
 (b) The case with $\delta a_s=2\,a_0$ in detail. 
 Density plots (color scale weighted by each density distribution) show the SF configuration and the SS configuration corresponding to the blue and red dots, respectively.
}
 \label{fig1}
\end{figure}

A useful quantity for characterizing the formation process of the supersolid is represented by 
the \textit{inverse participation ratio} (IPR), that measures the degree of localization (high IPR) or spread (low IPR) of a certain quantum state. For a continuous system, it can be defined as $IPR=\int |\psi|^{4}dV$. In the present case, it turns out to be proportional to the contact interaction energy $E_{\text{int}}$ [see Eq. \eqref{eq:GPenergy}],
\begin{equation}
IPR\propto E_{\text{int}}(t) = \frac{g}{2} \int n^2(\bm{r},t) d\bm{r}.
\label{eq:ipr}
\end{equation}
In order to facilitate the ongoing discussion, it is convenient to normalize the above expression as
\begin{equation}
\bar{E}_{\text{int}}(t)\equiv E_{\text{int}}(t)/E_{\text{int}}(0),
\label{eq:ren_ipr}
\end{equation}
which is also equivalent to normalizing the inverse participation ratio to its initial value at $t=0$.
The behavior of this quantity as a function of the time $t$ elapsed after the quench is shown in Fig. \ref{fig1}a, for different values of $\delta a_s$. It is noteworthy that all the cases presented exhibit the same qualitative behavior, which is exemplified in Fig. \ref{fig1}b for the case $\delta a_s=2.0\,a_0$.

Just after the quench, the system retains its initial character of a SF condensate for a certain period of time, during which it undergoes mostly breathing-like oscillations. 
These oscillations exhibit a consistent pattern among the different values of $\delta a_{s}$, as indicated by the blue arrows in Fig. \ref{fig1}a. After some time, the supersolid structure emerges, following a sudden increase of $\bar{E}_{\text{int}}$. As we shall see later on, this is a typical signature of an underlying formation process driven by the roton instability, namely the exponential growth of the corresponding mode, which becomes unstable. In Sec. \ref{sec:1Dmodel}, we will provide a detailed quantitative analysis and characterization of this behavior.

It should be emphasized that the SS state generated by this mechanism 
can display significant deviations from the supersolid ground state at the same value of $a_s$, due to the highly out-of-equilibrium nature of both the quench and the formation process (see also the discussion in Refs. \cite{Ch18,Ch19,wenzel2017}). 
In addition, since dissipation processes are not included in the present analysis, the system remains in such a highly excited state even long after the formation of the supersolid. 
However, it is worth noticing that for sufficiently low values of $\delta a_s$ (up to $\delta a_s=2\,a_0$, for the cases shown in Fig. \ref{fig1}a), the value $\bar{E}_{\text{int}}^*$ at which the instability saturates almost coincides with the equilibrium value $\bar{E}_{\text{int}}^{SS}$ for the ground state, indicated by  horizontal lines in the figure. This is no longer the case at higher values of $\delta a_s$, for which IPR saturation takes place at lower values than IPR for the SS ground state, $\bar{E}_{\text{int}}^*<\bar{E}_{\text{int}}^{SS}$. This effect becomes more prominent by increasing $\delta a_{s}$, owing to the increase of nonlinear effects by moving deeper into the SS phase (having started in all cases in the same initial configuration). 

In Fig. \ref{fig1}b, the time frame during which the system exhibits the instability is illustrated as a (blue) shaded area in between two consecutive oscillation maxima represented by blue and red dots. The density distributions of the corresponding states are shown in the insets. The first one corresponds to a state that can be still associated with the SF phase, despite displaying a weak density modulation. The other clearly corresponds to a well-formed SS state. 
Considering the above scenario, the time it takes for the supersolid to form from the instant of the quench, denoted as $\tau_{SS}$ and named \textit{formation time} in what follows, can be therefore conveniently defined by referring to the position of the red dot(s), as illustrated in the same figure.
Figure \ref{fig1}a shows that this formation time gets reduced by increasing $\delta a_s$. 

\section{Effective 1D description}
\label{sec:1Dmodel}

To gain a quantitative understanding of the superfluid formation process, we will employ the effective 1D model for an elongated dipolar condensate described in Ref. \cite{blakie2020variational}, simplifying the subsequent analysis. 
Owing to the strong transverse confinement of the present configuration, $\omega_{x}/\omega_{y,z}\approx0.15$, and the small asymmetry between the transverse trapping frequencies, $\omega_{y}/\omega_{z}\approx1$ (see Sec. \ref{sec:system}), this approximation is expected to be reasonably accurate. 

In summary, this approach consists in factorizing the condensate wave function as $\psi(\bm{r})=\varphi(x)\chi(y,z)$, where the transverse wave function $\chi(y,z)$ is conveniently taken as a Gaussian, 
\begin{equation}
\chi(y,z)=(\sqrt{\pi}l)^{-1}e^{-(\eta y^{2} + z^{2})/2\eta l^{2}},
\end{equation}
with $l=\sqrt{l_{y}l_{z}}$, $l_{y}$ ($l_{z}$) being the $1/e$ half width of the transverse density along the $y$-axis ($z$-axis), and $\eta=l_{y}/l_{z}$ the transverse anisotropy of the density distribution. From a Gaussian fit of the transverse profile of the initial SF density distribution we obtain $l_{y}\simeq 0.82\mu$m and $l_{z}\simeq 2.19\mu$m.

Integrating out the transverse directions, one obtains an effective 1D GP model where the contact interaction strength and the LHY coefficient get renormalized as $g_{1D}=g/(2\pi l_{y} l_{z})$, $\gamma_{LHY}^{(1D)}=\gamma_{\perp}\gamma_{LHY}$, with $\gamma_{\perp}=2/5\pi^{3/2}l^{3}$. The dipole-dipole effective interaction potential can be conveniently approximated in  momentum space by 
\begin{equation}
 \tilde{V}_{\text{dd}}^{(1D)}(q)=
 \frac{1}{1+\eta}\frac{\mu_0\mu_{Dy}^2}{2\pi l_{y} l_{z}}\left[q^2 e^{q^2}\text{Ei}(-q^2) +\frac{2-\eta}{3}\right],
 \label{eq:vdd}
\end{equation}
with $q\equiv \eta^{1/4}l k/\sqrt{2}$ and $k$ representing the momentum component relative to the $x-$axis.

The description can be further simplified by considering small deviations from uniformity, with linear density $n_{0}$. The longitudinal wave function can be expressed as $\varphi(x,t) =\sqrt{n_{0}} + \delta\varphi(x,t)$, with the latter term representing a small perturbation over the initial state $\varphi_{0}=\sqrt{n_{0}}$. The collective excitations of the condensate are then described by the associated Bogoliubov-de Gennes equations, see, e.g., \cite{ronen2006,Ch18,Ro19,blakie2020variational}. 
By looking for solutions of the form $\delta\varphi(x,t)\propto u(x)e^{-i\omega t} + v^{*}(x)e^{i\omega t}$ and considering that for a uniform system the excitations are plane waves of momentum $\hbar k$, namely $u_k(x)=U_k e^{-ikx}$ and $v_k(x)=V_k e^{-ikx}$, one finally obtains the following expression for the excitation energies \cite{blakie2020variational}
\begin{equation}
\label{eq:dispersion}
 \hbar\omega_k=\pm\sqrt{\frac{\hbar^2 k^2}{2m}\left(\frac{\hbar^2 k^2}{2m}+2n_0\tilde{V}(k) + 3\gamma_{LHY}^{(1D)}n_{0}^{3/2}\right)},
\end{equation}
where $\tilde{V}(k)$ denotes the Fourier transform of the interaction potential, see Eq. (\ref{eq:vdd}),
\begin{equation}
 \tilde{V}(k) = g_{1D} +\tilde{V}_{\text{dd}}^{(1D)}(\eta^{1/4}l k/\sqrt{2}).
\end{equation}

The above spectrum is characterized by a roton excitation (that is, a local minimum in the excitation dispersion relation) that softens to zero energy and becomes dynamically unstable when the $s-$wave scattering length is tuned below a certain critical value $a_{s}^{c}$ \cite{Ch18,blakie2020variational}.

\section{Stability analysis}
\label{sec:stability}

The quasi-1D effective formulation presented above allows  a straightforward stability analysis of our system after the quench. In order to do so, we need to establish a criterion to account for the non-uniformity of the system. 
In particular, we use the fact that for a continuous transition, the critical point for the SF-SS transition is expected to coincide with the roton instability, see Refs. \cite{Bl20,smith2023}.
Specifically, by defining $n(x)\equiv\int n(\bm{r})dydz$, we set $n_{0}=c n(0)$, where $c$ is chosen to reproduce the critical value of the $s-$wave scattering length $a_{s}^{c}=94.4a_0$ obtained from numerical simulations (see Sec. \ref{sec:protocol}). We find $c\simeq0.5$.

\begin{figure}[t]
 \centerline{\includegraphics[width=0.9\linewidth]{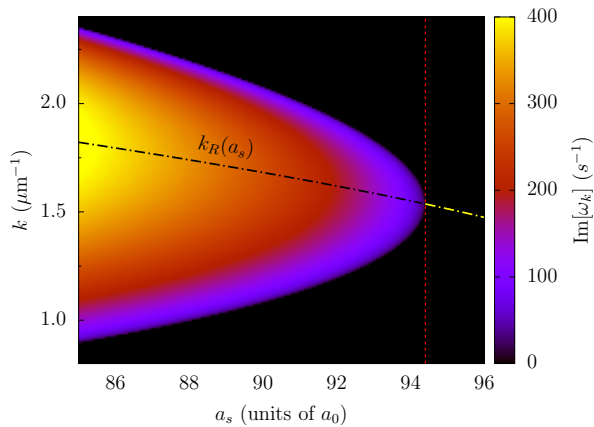}}
 \caption{
 Imaginary part of $\omega_k$, as a function of the $s-$wave scattering length $a_{s}$ and of the excitation momenta $k$. The vertical (red) dashed line indicates the critical value of the scattering length, $a_s^c=94.4a_0$. The black area corresponds to $\textrm{Im}[\omega_{k}(a_{s})]=0$. The (yellow) dot-dashed line represents the position of the roton mode, which disappears at $a_s^c$, where it is replaced by the most unstable mode (black dot-dashed line). The whole line is indicated as $k_R(a_s)$.}
 \label{fig2}
\end{figure}
In Fig. \ref{fig2} we show the imaginary part of $\omega_k$, as a function of the $s-$wave scattering length $a_{s}$ and of the excitation momenta $k$. We recall that the presence of imaginary frequencies in the spectrum is associated with exponentially growing modes that make the system modulationally unstable, if they are initially populated. In the figure, the vertical (red) dashed line corresponds to the critical value of the scattering length, $a_s^c=94.4a_0$. Above $a_s^c$ the spectrum is purely real, corresponding to a stable SF state. Below $a_s^c$ the frequency of some modes becomes imaginary, as indicated by the colored area. This unstable region broadens by decreasing the value of $a_{s}$. The dot-dashed line corresponds to the position of the most unstable mode, for $a_{s}<a_{s}^{c}$, which connects to that of the roton mode, for $a_{s}>a_{s}^{c}$. We indicate its position as $k_R(a_s)$. It is also worth noting that the LHY correction in Eq. \eqref{eq:dispersion} has a negligible contribution to the spectrum in Fig. \ref{fig2} for the parameter values at hand. 

\subsection{Supersolid formation}

Now that we have determined the properties of the excitation spectrum, we can revisit the dynamical behavior of the condensate after the quench, discussed in Sec. \ref{sec:protocol}, and examine the formation of the supersolid in terms of the emergence of exponentially growing modes. In particular, we use the fact that in the present case the exponentially growing modes take the form $u_k(x,t)=U_k e^{-ikx}e^{t/\tau_{k}}$, with  $\tau_{k}^{-1}\equiv\textrm{Im}[\omega_{k}]$ indicating the growth rate. We also make the assumption that instability is dominated by the most unstable mode, $k\to R$. 
Then, the normalized inverse participation ratio in Eqs. (\ref{eq:ipr}), (\ref{eq:ren_ipr}) can be approximated by 
\begin{align}
\bar{E}_{\text{int}}(t)&\simeq n_0^{-2}\int|\sqrt{n_0} + u_{R}(x,t)|^4dx
\nonumber\\
&=1 + 4 A e^{2t/\tau} + A^{2} e^{4t/\tau}\equiv 1 + f_{A,\tau}(t),
\label{eq:fit}
\end{align}
where we have defined $A\equiv n_0^{-1} U_{R}^{2}$ \footnote{The excitation momentum $k$ introduces a characteristic periodicity scale $\pi/k$. The integral can be written as an integer multiple of the unit integral over a single period, $\int dx=\ell\times\int_{0}^{\pi/k}dx$, with the total length of the system being $L=\ell\times\pi/k$. The final results depends on $L$ only through the  linear density $n_{0}$. }. The above expression, which depends on the two independent parameters $A$ and $\tau$, fixes the scaling behavior of the instability. In this respect, it is important to mention that while $\tau$ is an intrinsic property of the system, determined by Eq. \eqref{eq:dispersion}, $A$ depends through $U_k$ on the initial population of the unstable modes, namely on the preparation of the initial state. 

\begin{figure}[t]
 \centerline{\includegraphics[width=0.9\linewidth]{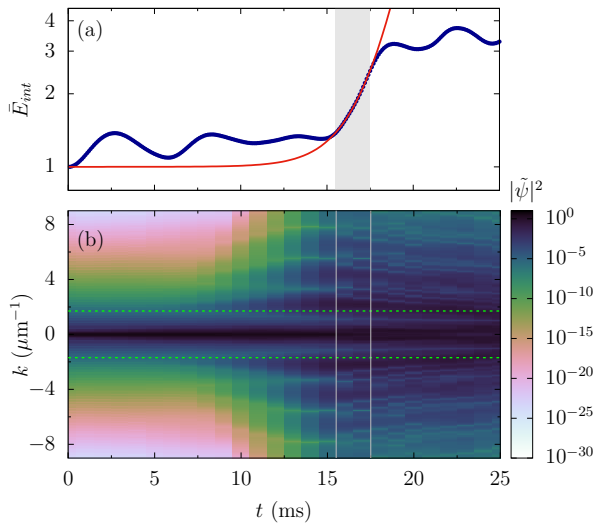}}
 \caption{Time evolution of the system after a quench with $\delta a_s=5\,a_0$.
 (a) Plot of $\bar{E}_{\text{int}}(t)$ (thick line) along with the fitting function in Eq. \eqref{eq:fit} (thin red line). 
 The grey area indicates the range in which the fit has been applied (see text).
 (b) Density plot of the momentum distribution along the $x-$direction as a function of time. 
 The horizontal (green) dashed line corresponds to the nominal most unstable modes at $\pm k_R$, see Fig. \ref{fig2}.
 Side peaks at integer multiples of the former are also visible (notice that the plot is in logarithmic scale).
 }
 \label{fig3}
\end{figure}
To illustrate the instability behavior, in Fig. \ref{fig3} we consider the case $\delta a_{s}=5\,a_{0}$, as an example. In the top panel (a), the evolution of $\bar{E}_{\text{int}}(t)$ (see also Fig. \ref{fig1}) is compared with the fitting function $1 +f_{A,\tau}(t)$ in Eq. \eqref{eq:fit}, using $A$ and $\tau$ as fitting parameters. The fit is restricted to the shaded area, where the instability becomes manifest.
Obviously, the simplified model in Eq. \eqref{eq:fit} cannot describe the initial fluctuations in the SF phase, that stem from the non uniform nature of the actual system, nor the collective oscillations of the SS state
, that clearly fall outside the regime of linear excitation of the initial state. However, it provides a clear explanation of the two different regimes observed after the quench, before the formation of the supersolid. The initial regime, in which the condensate retains its SF character, corresponds to the one in which the population of the unstable modes remains negligible, namely $f_{A,\tau}(t)\ll1$. 
Then, an evident exponential behavior emerges, leading to the formation of the supersolid, at $t=\tau_{SS}$. At this point, the system has already left the linear excitation regime, and Eq. \eqref{eq:fit} no longer applies. From $\tau_{SS}$ on, the dynamics of the system once again become predominantly driven by nonlinear effects, with quantum fluctuations being instrumental for stabilizing the system.

The growth of unstable modes is clearly visible in Fig. \ref{fig3}b, where we show a density plot representing the time evolution of the momentum distribution along the $x-$direction (see also Ref. \cite{Ch18}). The horizontal (green) dashed lines correspond to the nominally most unstable mode in Fig. \ref{fig2}, namely the roton excitations at $\pm k_R$. Notably, it is in good agreement with the result of the numerical simulation (see also Ref. \cite{Ch18,Ta19}). 

\begin{figure}
 \centerline{\includegraphics[width=0.9\linewidth]{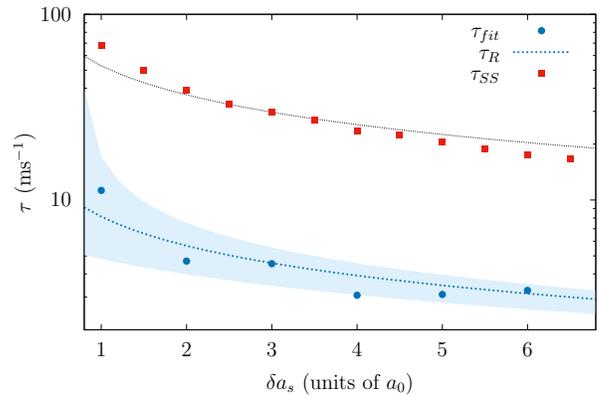}}
 \caption{Plot of the different characteristic times $\tau$ entering the formation of the supersolid. $\tau_{fit}$ is the inverse growth rate of the instability extracted from a fit of $\bar{E}_{\text{int}}(t)$ as in Fig. \ref{fig3}, to be compared with $\tau_{R}$ obtained from the stability analysis in Fig. \ref{fig2}. The light shaded area represents the variation of the latter upon a variation of $+10\%$ (downwards) and $5\%$ (upwards) of the linear density $n_{0}$. $\tau_{SS}$ is the formation time of the SS state defined in Fig. \ref{fig1}. The black dotted line corresponds to $\alpha\tau_R(\delta a_{s})$, with $\alpha\simeq6.5$ (see text).}
\label{fig4}
\end{figure}
The same analysis can be repeated for the other values of $\delta a_{s}$ considered in Sec. \ref{sec:protocol}.
In Fig. \ref{fig4} we compare the inverse growth rate extracted from the fit, $\tau_{fit}$, with the value $\tau_{R}\equiv\textrm{Im}[\omega_{R}]^{-1}$ corresponding to the most unstable mode in Fig. \ref{fig2}.
We also show, as a light shaded area, how $\tau_{R}$ changes by varying the linear density $n_{0}$.
As one may expect, the instability mechanism is more effective for larger densities. This observation is consistent with the results of the GP simulations, which show that the formation of the SS pattern first occurs at the center of the condensate, where the density is highest, and subsequently spreads outward toward the lower density region in the tail. Overall, the agreement between the model predictions and the numerical data is remarkably good, indicating that the model captures the essential physics of the system despite the inherent simplifications leading to Eq. \eqref{eq:dispersion}, such as the assumption of uniformity.

In Fig. \ref{fig4} we also show the total formation time $\tau_{SS}$, as red squares. Interestingly, the dependence of $\tau_{SS}$ on $\delta a_{s}$ follows a behavior similar to that of $\tau_{R}$, namely $\tau_{SS}\simeq \alpha\tau_R(\delta a_{s})$ (notice the log scale of the vertical axis), with $\alpha$ being a proportionality factor. In the present case we find $\alpha\simeq6.5$, see the black dotted line in the figure. 
Actually, the figure shows that there must be also a slight correction to the scaling, namely $\alpha=\alpha_{\delta a_{s}}$.
This can be explained as follows.
From the simplified model in Eq. \eqref{eq:fit} it is straightforward to express the supersolid formation time as
\begin{equation}
 \tau_{SS} = \tau_R \left[\ln\left(\frac{\sqrt{n_0}}{U_{R}}\right)
 +\frac12\ln\left(\sqrt{3 + \bar{E}_{\text{int}}^*}-2\right)
 \right].
\end{equation}
This result confirms that the scaling of $\tau_{SS}$ as a function of $\delta a_{s}$ is indeed mostly determined by the behavior of $\tau_R(\delta a_{s})$, with $\bar{E}_{\text{int}}^*(\delta a_{s})$ and $U_R(\delta a_{s})$ providing a logarithmic correction. Regarding the initial population of the unstable modes ($U_R$), it is worth noting that -- given a certain realization of the initial state (and of its momentum distribution) -- it may slightly depend on $a_{s}$ due to the varying position of the roton momentum $k_R(a_{s})$ (see Fig. \ref{fig3}). Although we do not include temperature in our considerations, we signal that if there is an increase in the initial population at the roton mode because of thermal excitation then the formation time of the supersolid would be shortened with respect to the $T=0$ prediction, as has been observed in experiments \cite{Bo19,biagioni2022,alana2022crossing}. It is also important to remark that the logarithmic terms, though they only provide a correction to the scaling, are essential for fixing the value of the proportionality constant $\alpha_{\delta a_s}$. In particular, we find that the major contribution comes from $\ln(\sqrt{n_0}/U_{R})$ in the range of values of $\delta a_s$ considered here.

\section{Conclusions}
\label{sec:conclusions}

We have characterized the role of the roton instability in the formation of a supersolid state of an elongated dipolar condensate across a continuous superfluid-supersolid transition, as experimentally reported by Biagioni \textit{et al.} \cite{biagioni2022} and theoretically investigated in Ref. \cite{alana2022crossing}. 
By means of numerical simulations based on the extended Gross-Pitaevskii equation, we have systematically analyzed the dynamic behavior of the system after a quench of the contact interactions considering various amplitudes $\delta a_s$ of the interaction quench. We have also computed the corresponding excitation spectrum by using the effective one-dimensional approach of Blakie \textit{et al.} \cite{blakie2020variational}, finding that the calculated growth rates of the unstable roton modes provide an accurate characterization of the behavior observed in the simulations.

Our main finding is that the scaling behavior of the formation time $\tau_{SS}$ of a supersolid state as a function of $\delta a_s$ is mostly determined by the scaling of the inverse growth rate of the roton mode, namely $\tau_{SS}(\delta a_{s})\simeq \alpha_{\delta a_{s}}\tau_R(\delta a_{s})$, with $\alpha_{\delta a_{s}}$ being a proportionality factor fixed by the initial population of the (most) unstable mode. Remarkably, this implies that the supersolid formation time $\tau_{SS}$ is determined by the properties of the \textit{initial} state rather then by the energy difference between the initial superfluid state and the final supersolid state, as it was speculated in Refs. \cite{Bo19,alana2022crossing}.
The present analysis -- that admits a straightforward extension also to other cases besides a quench (e.g., by means of a time rescaling as discussed in Ref. \cite{Ch18})
-- offers a deeper insight into the formation dynamics of the supersolid structures in dipolar condensates, thereby providing a valuable framework for future experiments and theoretical studies.

\begin{acknowledgments}
This work was supported by Grant PID2021-126273NB-I00 funded by MCIN/AEI/ 10.13039/501100011033 and by ``ERDF A way of making Europe'', by the Basque Government through Grant No. IT1470-22, and by the European Research Council through the Advanced Grant ``Supersolids'' (No. 101055319).
\end{acknowledgments}

\end{document}